\documentclass{jpp}
\usepackage{graphicx}

\usepackage{hyperref}
\usepackage[utf8]{inputenc}
\usepackage[T1]{fontenc}
\usepackage{amsmath}
\usepackage{physics}

\usepackage{physics}
\usepackage{xcolor}
\usepackage{comment}
\usepackage{float}

\newcommand{\e}{\epsilon}
\newcommand{\ky}{k_y}
\newcommand{\kz}{k_z}
\newcommand{\ko}{k_0}

\newcommand{\st}{\textit{s.t. }}

\newtheorem{theorem}{Theorem}[section]

\usepackage{layouts}

\shorttitle{Analytic power loss at the lower hybrid resonance in ANTITER IV}
\shortauthor{V. Maquet et al}

\title{Analytical edge power loss at the lower hybrid resonance: comparison with ANTITER IV and application to ICRH systems}

\author{V. Maquet\aff{1,2}
  \corresp{\email{Vincent.Maquet@ulb.be}},
  A. Druart\aff{2,3},
  A. Messiaen\aff{1}}

\affiliation{\aff{1}Laboratory for Plasma Physics - ERM/KMS, Avenue de la Renaissance 30, B-1000 Brussels.
\aff{2}Université Libre de Bruxelles,
B-1050 Brussels.
\aff{3}International Solvay Institutes, CP 231, B-1050 Brussels.}

\begin{document}

\maketitle

\begin{abstract}
In non-inverted heating scenarios, a lower hybrid (LH) resonance can appear in the plasma edge of tokamaks. This resonance can lead to large edge power deposition when heating in the ion cyclotron resonance frequency (ICRF) range. In this paper, the edge power loss associated with this LH resonance is analytically computed for a cold plasma description using an asymptotic approach and analytical continuation. This power loss can be directly linked to the local radial electric field and is then compared to the corresponding power loss computed with the semi-analytical code ANTITER IV. This method offers the possibility to check the precision of the numerical integration made in ANTITER IV and gives insights in the physics underlying the edge power absorption. Finally, solutions to minimize this edge power absorption are investigated and applied to the case of ITER's ion cyclotron resonance heating (ICRH) launcher. This study is also of direct relevance to DEMO.
\end{abstract}

\section{Introduction}
A potentially important power loss mechanism for ion cyclotron resonance heating (ICRH) arising in the presence of a lower hybrid (LH) resonance in the edge of a tokamak plasma has recently been discussed in \cite{Messiaen_2020}. The possibility of this power loss was already pointed out in earlier work (\cite{morales,Lawson_1992}) and can be linked to a confluence between the fast and the slow wave at the LH resonance in non-inverted heating scenarios (\textit{i.e.} where $\omega>\omega_{ci}$ where $\omega_{ci}$ is the the cyclotron frequency of the majority ions and $\omega$ is the driving angular frequency of the antenna) for toroidal wavenumber $k_z$ smaller than the propagation constant in vacuum $k_0$. The same paper provides simple rules to minimize the power loss at this LH resonance constraining the current distribution on the strap array in amplitude and phase. 

Edge power loss in tokamaks should be avoided as it can lead to a reduction of the deposited heating power in the plasma core and to deleterious impurity release from the first wall of the device. A correlation between ICRH related impurity release and low $\abs{k_z}<k_0$ present in the $k_z$ spectrum launched by ICRH antennas was investigated in \cite{maquet_messiaen_2020}. This paper also proposes a new non-conventional antenna strap phasing minimizing power losses in the edge of the tokamak for effective heating of the core plasma.
\newline

In the edge of a tokamak, the plasma can be approximated by the cold plasma dispersion description. A recent upgrade of ANTITER II, called ANTITER IV, is now describing the waves launched by an ICRH antenna in the cold plasma limit including the full description of the fast and slow waves confluence and the LH resonance aspects (\cite{messiaen_2021}). The present paper analytically derives the power loss at the LH resonance in section \ref{sec:1}, compares the results with the power loss computed numerically by ANTITER IV in section \ref{sec:2} and applies the results to relevant operational scenarios for the ITER ICRH antenna in section \ref{sec:3}.

\section{Analytical derivation}
\label{sec:1}
The edge plasma is described by the cold dielectric plasma tensor and Maxwell's equations expressed in the radial $x$ direction for a slab geometry and Fourier analysis in the $y,z$ directions where $z$ represents the direction along the total steady magnetic field $B_0$. The plasma wave model considered leads to a system of 4 first order ordinary differential equations (ODEs) of the form $Y(x)'=A(x) Y(x)$:
{\small
\begin{align}
    \dv{}{x}\mqty(i\omega B_z\\ E_y \\ i\omega B_y\\ E_z) = \frac{1}{\epsilon_1}\mqty(-\ky\e_2 & \ko^2 \e_2^2 +(\e_1\kz^2-\ko^2\e_1^2)& \e_2 \kz & -\ky\kz\e_1 \\
    \e_1-\frac{\ky^2}{\ko^2} & \ky\e_2 & \frac{\ky\kz}{\ko^2} & 0 \\
    0 & \ky\kz\e_1 & 0 & \e_1 (\ko^2\e_3-\ky^2) \\
    -\frac{\ky\kz}{\ko^2} & \kz\e_2 & \frac{\kz^2}{\ko^2}-\e_1 & 0)\mqty(i\omega B_z\\ E_y \\ i\omega B_y\\ E_z).
    \label{eq:model}
\end{align}
}
Moreover,
\begin{align}
    E_x &= -\frac{k_y}{\ko^2 \e_1} \omega B_z + \frac{k_z}{\ko^2 \e_1}\omega B_y
    -i\frac{\e_2}{\e_1} E_y,    \label{eq:Ex} \\
    \omega B_x &= k_y E_z - k_z E_y.
\end{align}
In the expressions above, $\e_1$, $\e_2$ and $\e_3$ are the components of the cold dielectric plasma tensor (\cite{swanson2012plasma}). The field components $E_y$ and $B_z$ can be associated to the fast wave components of interest for ICRH and $E_z$ and $B_y$ can be associated to the slow wave components. This system is singular at the location $x_0$ where $\e_1(x_0)=0$ which corresponds to the LH resonance in the cold plasma description. In what follows, the method used to derive the power loss at this LH resonance is similar to the one in \cite{Faulconer_1994} where the power loss at the Alfvén resonance was obtained from an asymptotic expansion of the system of differential equations and from an analytical continuation around the singularity.

\subsection{Asymptotic expansion in the vicinity of the LH resonance}
\noindent
Choosing the position of the LH resonance at the origin $x=0$, one can expand $\e_1(x)$ in a Taylor series as $\epsilon_1(x) =\epsilon_1'(0)x+\order{x^2}$ where $\e_1'$ is the derivative of $\e_1$ with respect to $x$. This leads to an asymptotic expression of $A(x)$ of the system \eqref{eq:model}:
\begin{align}
    A(x) = \frac{A_0}{\e_1' x} + \order{x^0}
    \label{eq:expansion}
\end{align}
with $\e_1'\equiv\e_1'(0)$ and where
\begin{align}
    A_0 \equiv \mqty(-\ky\e_2 & \ko^2\e_2^2 & \kz \e_2 & 0 \\
    -\frac{\ky^2}{\ko^2} & \ky \e_2 & \frac{\ky\kz}{\ko^2} & 0\\
    0 & 0 & 0 & 0 \\ 
    -\frac{\ky\kz}{\ko^2} & \kz\e_2 & \frac{\kz^2}{\ko^2} & 0).
\end{align}
A straightforward computation shows that the matrix $A_0$ satisfies
\begin{equation}
    A_0^2=0\label{nihil}.
\end{equation}
This property will be fundamental in the forthcoming computations. The matrix $A_0$ can also be expressed in a way that will become handy later: 
\begin{align}
    A_0=\frac{1}{\ko^2}\mqty(\ko^2\e_2\\\ky\\0\\\kz)\mqty(-\ky & \e_2\ko^2 & \kz & 0).
    \label{eq:trick}
\end{align}

\subsection{Fields near the resonance}
\noindent
To derive expressions for the tangential $(y,z)$ fields near the resonance we consider $x$ as being a \textit{complex} variable (\textit{i.e.}, we will consider the analytic continuation of our system of equations over the complex plane). Close to $x=0$, one can truncate the expansion \eqref{eq:expansion} as
\begin{align}
A(x)\simeq\frac{A_0}{\e'_1\,x}.
\end{align}
In this approximation, using the property \eqref{nihil}, we observe that
$A(x)$ and its primitive commute. This enable us to use the general theorem derived in Appendix \ref{app:thm}. The solution to the system \eqref{eq:model} reads
\begin{align}
    \mqty(i\omega B_z\\ E_y \\ i\omega B_y\\ E_z)&=\exp\qty(\int^x A(x')\dd x')\mathbf C
    =\exp\qty(\frac{A_0}{\e_1'} \log(x)) \mathbf C
    =\qty( I + \frac{A_0}{\e_1'} \log(x)) \mathbf C,
    \label{eq:sol}
\end{align}
where $\mathbf C\equiv\mqty(C_1,C_2,C_3,C_4)$ is a constant column vector depending on the initial conditions of the problem. Here, $\log$ denotes the principal value of the complex logarithm. The last equality in \eqref{eq:sol} is \textit{exact} due to the property \eqref{nihil}. More explicitly, our solution is given by
\begin{align}
    \mqty(i\omega B_z\\ E_y \\ i\omega B_y\\ E_z)= \mqty(C_1 \\ C_2 \\C_3 \\ C_4) + \alpha(x) \mqty(\ko^2\e_2\\ \ky\\ 0 \\ \kz)\mqty(-k_y C_1 &\e_2\,\ko^2\, C_2 &k_z C_3 &0),\label{eq:explicit_sol}
\end{align}
with $\alpha(x)\equiv\frac{\log(x)}{\ko^2\e_1'}$ and where we used relation \eqref{eq:trick}. 

The expression above clearly shows that all fields except $B_y$ are singular at the resonance. Remarkably, one can nevertheless construct a particular combination of them which remains \textit{non-singular} at $x=0$. Left multiplication of equation \eqref{eq:explicit_sol} by the line vector $\mqty(-\ky & \e_2\ko^2 & \ko & 0)$ leads to
\begin{align}
    \ko^2\e_2E_y-\ky i\omega B_z+\kz i\omega B_y = \ko^2\e_2 C_2 -\ky C_1+\kz C_3\equiv \Upsilon.
    \label{aff:1}
\end{align}
There is a clear relationship between this combination of the fields and the radial field $E_x$: recalling that the latter takes the form given in \eqref{eq:Ex}, $\Upsilon$ can be written as
\begin{align}
    \Upsilon=i\ko^2\e_1E_x.
\end{align}
We finally get from \eqref{eq:explicit_sol}
\begin{align}
        i\omega B_z &= C_1 + \alpha(x) \e_2 \ko^2 \Upsilon ,\\
        E_y &= C_2 +  \alpha(x) \ky \Upsilon,\\
        i\omega B_y &= C_3,\\
        E_z &= C_4 + \alpha(x) \kz \Upsilon.
        \label{eq:fields}
\end{align}
These are the expressions of the singular part (\textit{i.e.} up to $\order{x^0}$ corrections) of the $y,z$ fields in the vicinity of the LH resonance (located at $x=0$).

\subsection{Power loss at the LH resonance}
\noindent
Our next goal consists in computing the power loss at the LH resonance using the expressions of the fields derived above. In our case, the power loss $\Delta P$ is given by the difference of the Poynting flux $S(x)$ after and before the resonance:
\begin{align}
    \Delta P\equiv \lim_{\kappa\to 0^+}S(x)\eval^{x=+\kappa}_{x=-\kappa}.
\end{align}
The Poynting flux can be written as
\begin{align}
    S(x) &\equiv \Re(E_y H_z^* - E_z H_y^*) \\
    &=\frac{-1}{\omega \mu_0} \Im[E_y (i\omega B_z)^* - E_z (i\omega B_y)^*].
    \label{eq:poynt}
\end{align}
Defining $a\equiv k_yC_1$, $b\equiv k_0^2\e_2C_2$ and $c\equiv k_zC_3$, one has
\begin{align}
    \Im\qty[E_z (i\omega B_y)^*]
    &= \Im(C_4 C_3^*) + \Im[\alpha(x) c^* \Upsilon]\\
    \Im\qty[E_y (i\omega B_z)^*]
    &= \Im(C_2 C_1^*) + \Im[\alpha^*(x) b \Upsilon^* + \alpha(x) a^* \Upsilon].
\end{align}
Making use of these expressions and of the identity $\Upsilon=-a+b+c$, the Poynting flux can be rewritten as
\begin{align}
    S(x) &= \frac{-1}{\omega\mu_0}\qty{\Im(C_2 C_1^*-C_4 C_3^*)+\Im[\alpha^*(x) b \Upsilon^* + \alpha(x) a^* \Upsilon-\alpha(x) c^* \Upsilon]}\\
    &=\frac{-1}{\omega\mu_0}\Im(C_2 C_1^*-C_4 C_3^*)+\frac{\Im[\alpha(x)]}{\omega\mu_0}\abs{\Upsilon}^2.
\end{align}
We finally have
\begin{align}
    \Delta P =\lim_{\kappa\to 0^+} \eval{S(x)}_{x=-\kappa}^{x=+\kappa}=\frac{\abs{\Upsilon}^2}{\omega\mu_0}\qty(\lim_{\kappa\to 0^+} \eval{\Im[\alpha(x)]}_{x=-\kappa}^{x=+\kappa}).\label{eq:pre_DP}
\end{align}
On the other hand, one has the identity
\begin{align}
    \Im[\alpha(x)]\eval_{x=-\kappa}^{x=+\kappa}\simeq-\frac{\pi}{k_0^2\abs{\epsilon'_1}}.\label{eq:alpha}
\end{align}
A proof of this equality is given in Appendix \ref{app:alpha}.
Plugging \eqref{eq:alpha} into \eqref{eq:pre_DP} leads to the final result:
\begin{align}
    \Delta P &= \frac{-\pi}{\ko^2\omega\mu_0\abs{\e_1'}}\abs{\Upsilon}^2 \label{aff:2}\\
    &=  \frac{-\pi\ko^2}{\omega\mu_0\abs{\e_1'}} \abs{\e_1 E_x}^2.
    \label{aff:3}
\end{align}
This equation confirms that power is indeed lost crossing the LH resonance. Relation \eqref{aff:3} gives physical insight in the physics underlining the edge power loss: it does not depend on the toroidal electric field $E_z$ but on the local radial electric field $E_x$. The power loss also inversely depends on the derivative of the first dielectric tensor $\e_1$ component: a larger density gradient in the edge will lead to lower power losses. Moreover, as relation \eqref{aff:3} is exact, it provides an opportunity to verify the numerical integration done in ANTITER IV when crossing this LH resonance.

\subsection{Parallelism with previous works}
The result \eqref{aff:3} reminds of the work of \cite{Faulconer_1994}, where only the fast wave was taken into account in the plasma description. This leads to a system of 2 first order ODEs:
\begin{align}
    \dv{}{x}\mqty(i\omega B_z\\ E_y) = \frac{1}{u}\mqty(-\e_2\ko^2k_y & -u^2 +\e_2^2\ko^4 \\
    u-k_y^2 & \e_2\ko^2k_y) \mqty(i\omega B_z\\ E_y).
\end{align}
Here,
\begin{align}
    u\equiv\ko^2\e_1-k_z^2.
\end{align}
The authors found that the power lost at the Alfvén resonance ($u=0$) is proportional to the square of $\abs{u E_x}$:
\begin{align}
    \Delta P=\frac{-\pi}{\omega \mu_0 \abs{u_0'}}\abs{u E_x}^2.
    \label{eq:alfven}
\end{align}
A clear symmetry can be found between \eqref{eq:alfven} and \eqref{aff:3}.

\section{ANTITER IV}
\label{sec:2}
The previous results can now be used to estimate the accuracy of the calculation in ANTITER IV. ANTITER is a semi-analytic code describing an antenna in front of a plasma in plane geometry in the cold plasma limit. The code uses Fourier analysis in the previously defined $(y,z)$ directions and numerical integration in the radial $x$ one. An ideal Faraday screen is assumed at the antenna mouth together with single-pass absorption in the plasma (\cite{Messiaen_2010}). 

\begin{figure}
    \centering
    \includegraphics[width=8cm]{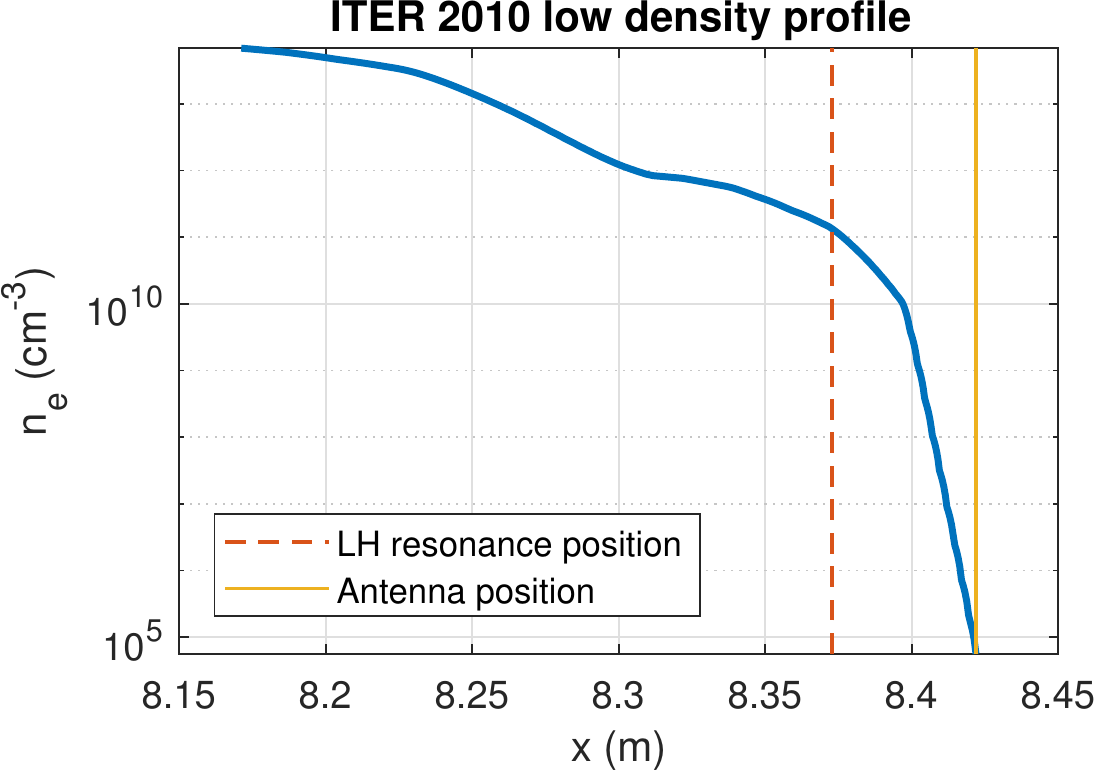}
    \caption{ITER 2010 low electron density profile. The antenna and the LH resonance position are also displayed.}
    \label{fig:denslow}
\end{figure}

The antenna is described by a set of boxes recessed into a metal wall containing infinitely thin straps. The edge plasma electron density profile used for the study is the ITER worst case plasma profile for IRCH (2010low -- \cite{15Xtreport}) as it is representative of the large SOL to be expected in large machines like ITER or DEMO. This electron density profile is presented in the figure \ref{fig:denslow} along with the antenna and the LH positions.

\begin{figure}
    \centering
    \includegraphics[width=12cm]{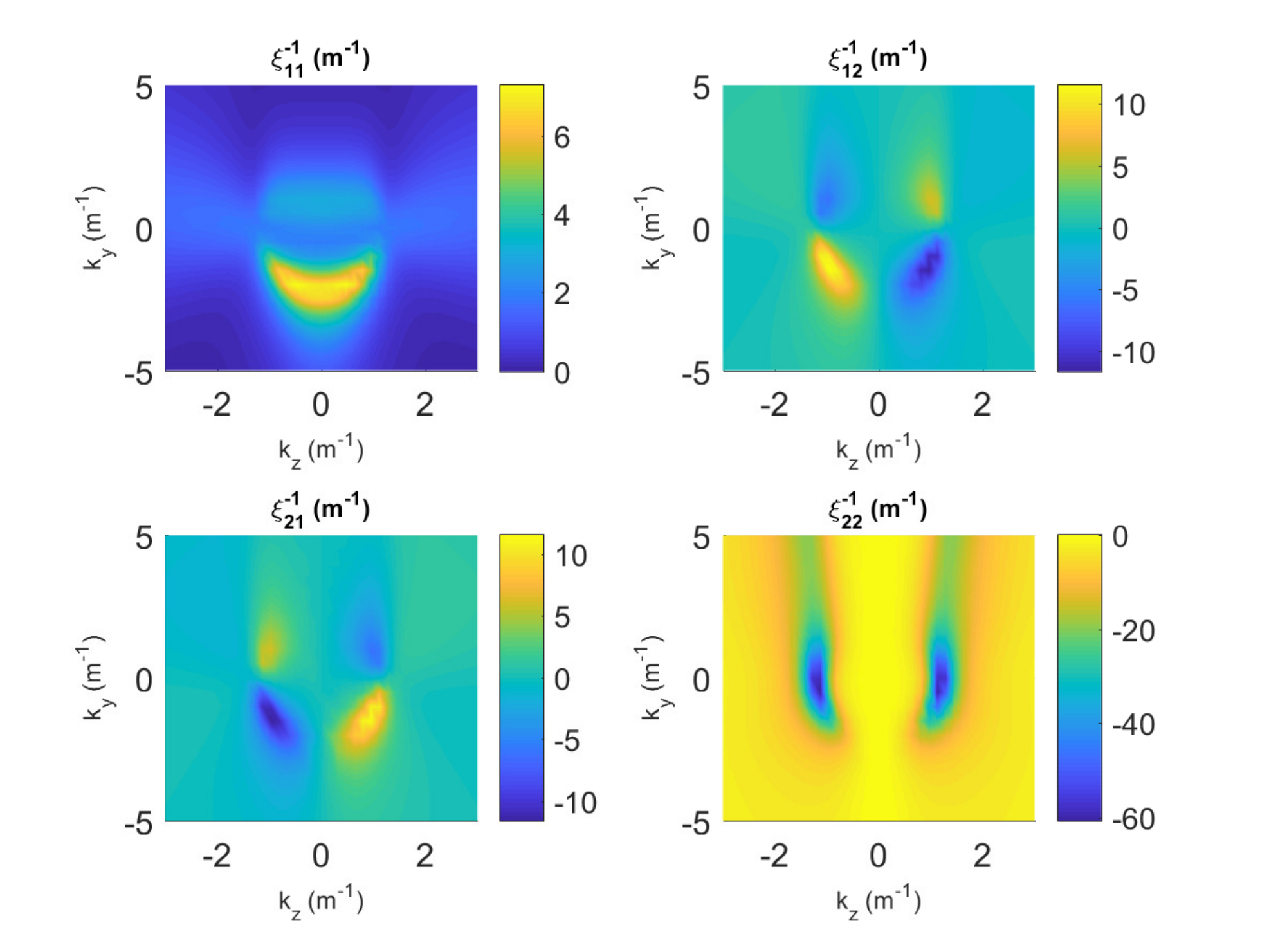}
    \caption{The four plasma impedance matrices seen at the antenna position as a function of $(k_y,k_z)$.}
    \label{fig:impedance_matrices}
\end{figure}

ANTITER II is only describing the fast wave, based on the fact that the fast and slow wave component can be considered decoupled in first approximation in the ion cyclotron range of frequencies for a large density range. This is no longer true at resonances. ANTITER IV extends the above description to include a detailed description of both the slow and fast waves (\cite{messiaen_2021}). The LH resonance is handled by adding a small amount of collisions in the cold plasma tensor which corresponds to adding an imaginary part to the dielectric tensor components. The plasma part is finally described at the antenna position by four admittance matrices
\begin{align}
    \mqty(\omega B_z\\ \omega B_y)=\mqty(\xi^{-1}_{11} & \xi^{-1}_{12} \\ \xi^{-1}_{21} & \xi^{-1}_{22})\mqty(E_y\\ E_z)
\end{align}
describing the relationship between the four tangential plasma components $E_y$, $E_z$, $B_z$ and $B_y$ for all the wavelets $(k_y,k_z)$ of the Fourier expansion. The real part of the four impedance matrices found at the antenna position (\textit{i.e.} at the lower end of the electron density profile of figure \ref{fig:denslow}) are presented in figure \ref{fig:impedance_matrices}. These matrices are important for the derivation of the active Poynting’s power flux.
\newline

\subsection{Fields near the LH resonance with ANTITER IV}
\begin{figure}
    \centering
    \includegraphics[width=\linewidth]{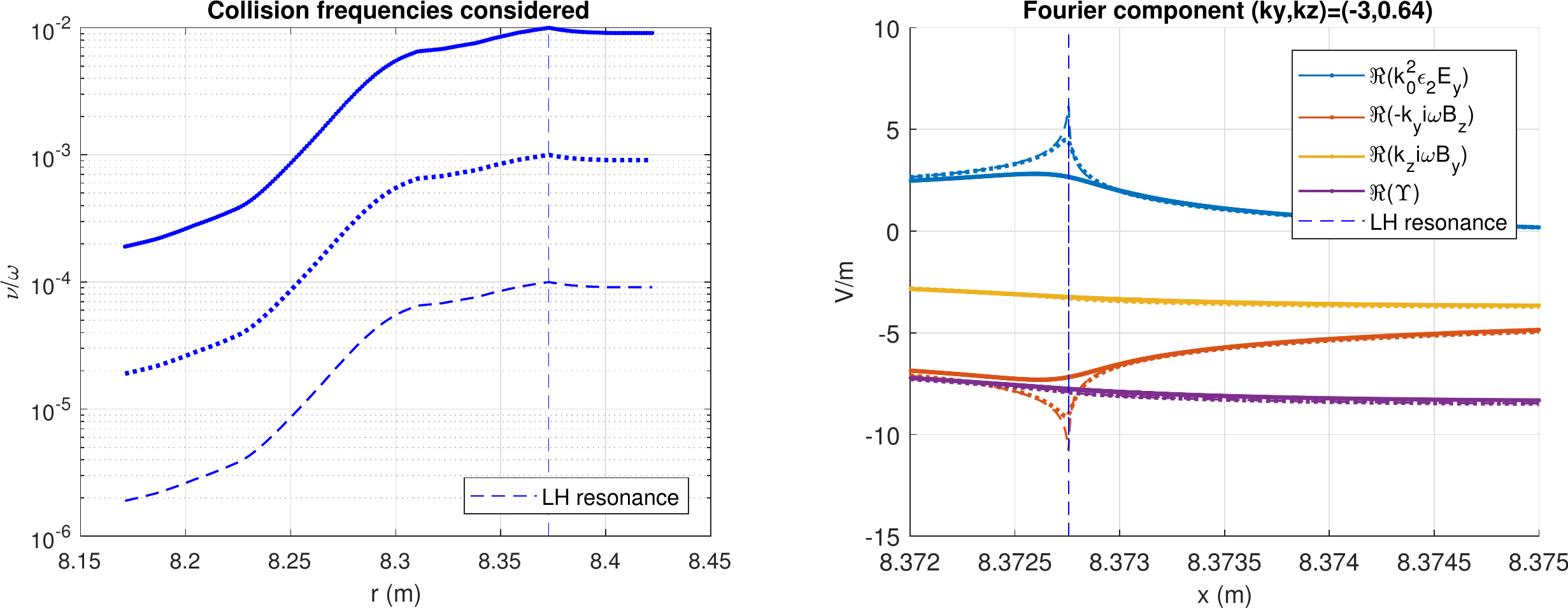}
    \caption{Illustration of the near singular behavior of individual components of the fields and their finite sum for a pair $(\ky,\kz)$ around the LH resonance.}
    \label{fig:1}
\end{figure}

We first investigate equation \eqref{aff:1}. If ANTITER IV correctly describes the waves at the LH resonance, relation \eqref{aff:1} should lead to a non-singular behaviour. The fields ultimately depend on the amount of collisions added to the dielectric tensor terms in order to integrate the system \eqref{eq:model}, in the vicinity of the singularity, by analytical continuation in the complex plane. The collision coefficient in ANTITER IV does not bear a physical meaning and its sole purpose is to bypass the resonance. Three collisions profile differing by one order of magnitude are presented in figure \ref{fig:1}a. Each component of the Fourier fields $\ko^2\e_2E_y$, $\ky i\omega B_z$, $\kz i\omega B_y$ and their sum are presented figure \ref{fig:1}b for the three collision coefficient selected. As expected from relation \eqref{eq:fields}, the fields $E_y$ and $B_z$ computed for low collision coefficient show near singular behaviour at the lower hybrid resonance while their sum $\Upsilon$ stays regular even for vanishing amounts of collisions. A smaller number of collisions leads to smaller integrating steps in ANTITER IV but improves the accuracy of the power loss calculation. Therefore, theoretical expressions like \eqref{aff:1} and \eqref{aff:3} gives an opportunity to assess the precision of the integration made in ANTITER IV. For the rest of the computations, the second collisions coefficient profile presented in figure \ref{fig:1}a is selected.

\subsection{Power loss at the LH resonance with ANTITER IV}
\begin{figure}
    \centering
    \includegraphics[width=\linewidth]{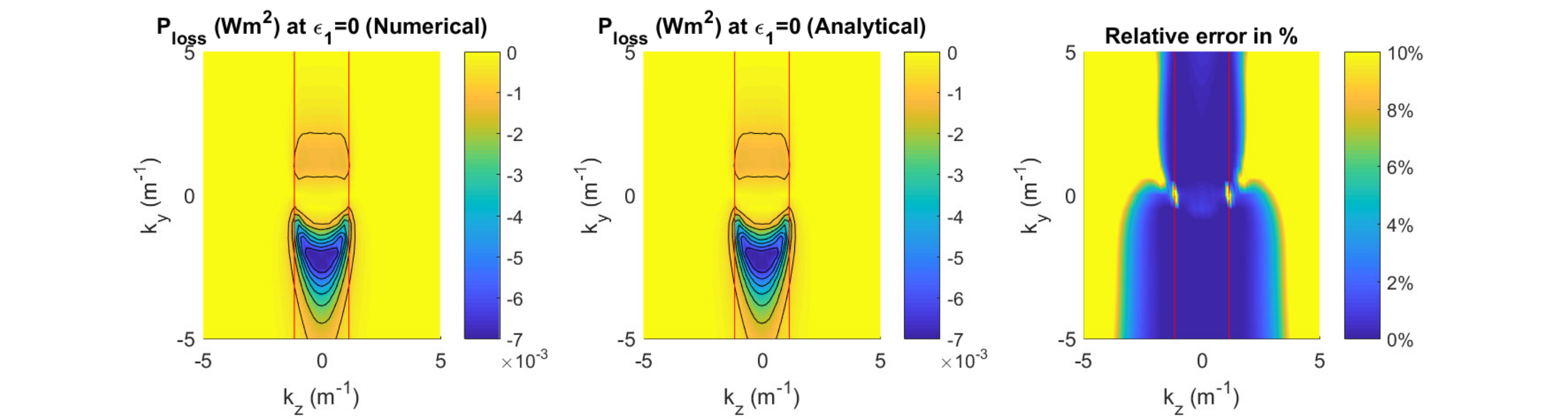}
    \caption{Power loss at the LH resonance for a pure $E_y(k_z,k_y)$ excitation. Red lines delimits the $\abs{k_z}<k_0$. Here $k_0=1.15$ m$^{-1}$.}
    \label{fig:Ey1}
\end{figure}

The power losses at the LH resonance using the Poynting flux \eqref{eq:poynt} and the analytical power loss \eqref{aff:3} computed with the fields of ANTITER IV are compared. This exercise is performed for a pure fast wave excitation (\textit{i.e.} $E_y(k_y,k_z)=1$ at the plasma edge while ensuring $E_z(k_y,k_z)=0$). For this specific excitation, the power loss is limited to the wavenumbers smaller than the wave propagation constant in vacuum  $\abs{k_z}<k_0$ as they correspond to the fast wave undergoing a wave confluence with the slow wave. This fact is verified in figure \ref{fig:Ey1}. We also observe that the relative error between the analytical Poynting flux given in \eqref{aff:3} and the numerical integration of ANTITER IV is negligible in the region of interest (\textit{i.e.} where the power loss is not negligible). The same test can be performed for a pure slow wave excitation (\textit{i.e.} $E_z(k_y,k_z)=1$  and $E_z(k_y,k_z)=0$). Here we see a strong interaction which is no more limited to the region $\abs{k_z}<k_0$ but extending to $k_0<k_z<2$ m$^{-1}$. A negligible relative error between the two methods is again observed.

\begin{figure}
    \centering
    \includegraphics[width=\linewidth]{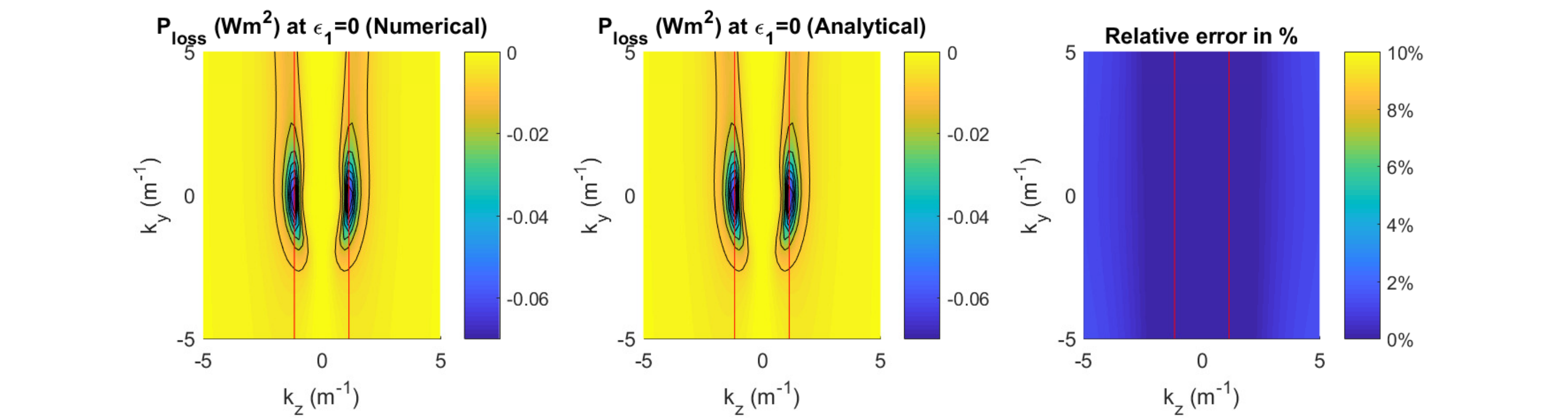}
    \caption{Power loss at the LH resonance for a pure $E_z(k_z,k_y)$ excitation. Red lines delimits the $\abs{k_z}<k_0$. Here $k_0=1.15$ m$^{-1}$.}
    \label{fig:Ez1}
\end{figure}

These results give further confidence in the ANTITER IV calculations. They hint at possibilities to minimize the edge power absorption. For a field-aligned antenna with a field-aligned Faraday shield (FS), figure \ref{fig:Ey1} shows that the power losses can be minimized by avoiding $\abs{k_z}<k_0$ in the power spectrum. The antenna power spectrum can be easily modified by shaping the $E_y$ spectrum excited by the antenna (\textit{i.e.} by varying the current amplitude and phase distribution over the straps). A FS that is not aligned with the background magnetic field will excite a spurious $E_z$ spectrum that can in turn lead to significant new losses for low $k_z$ above $\abs{k_z}<k_0$ as shown in figure \ref{fig:Ez1}. These new losses can be reduced by further depleting the low $k_z$ part of the power spectrum at the expense of a reduction in the power coupled to the plasma core. We also see that for an equal excitation of $E_y$ and $E_z$, the losses due to $E_z$ are one order of magnitude larger than the losses due to $E_y$.

Finally, one can also minimize power losses using gas puff (\cite{ZHANG2019364}) which will lead to larger density gradient near the lower hybrid location. This last method can lead to a substantial decrease in the edge power losses and at the same time increases the power coupling to the core plasma. 

\section{Application}
\label{sec:3}
The results of the previous sections indicate how to minimize power losses into the presence of a LH resonance in the plasma edge for a given plasma profile. Here we use ANTITER IV to minimize those losses for a given plasma density profile using a multidimensional minimization procedure.

\subsection{ITER-like antenna}
\begin{figure}
    \centering
    \includegraphics[width=\linewidth]{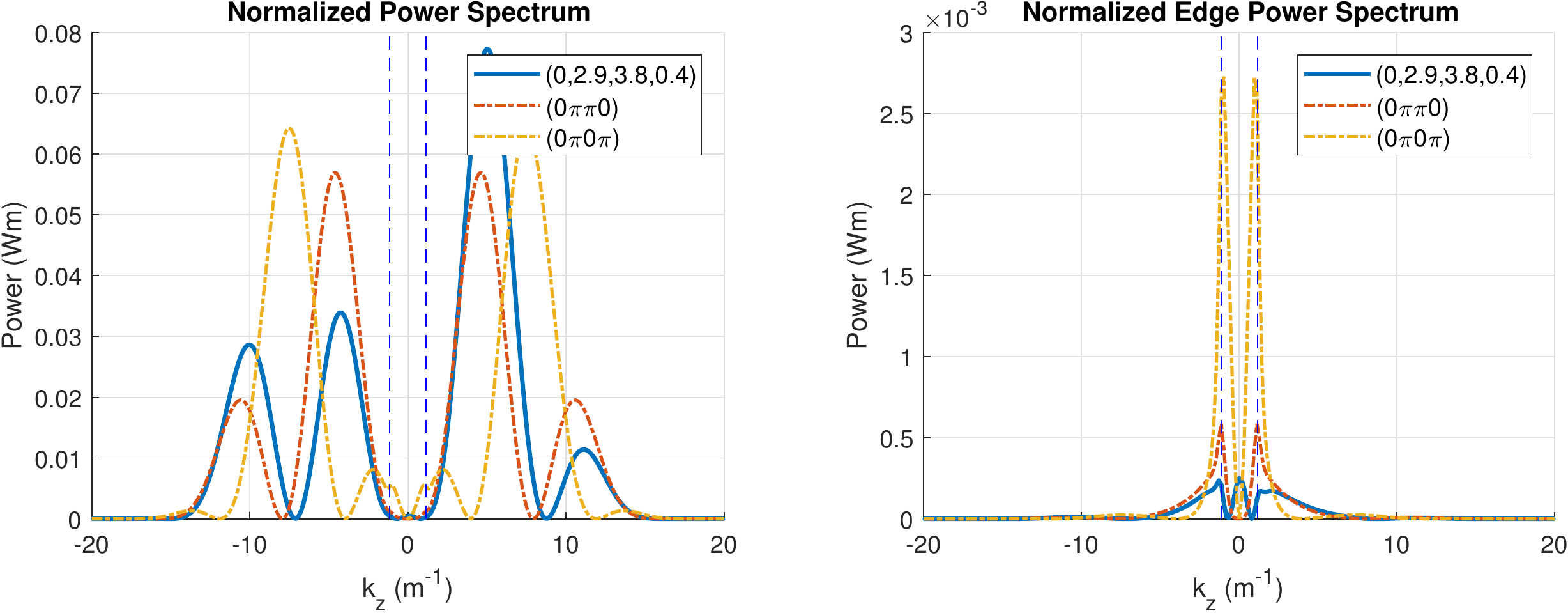}
    \caption{(a) Normalized $k_z$ power spectrum and (b) normalized $k_z$ edge power loss spectrum for a current distribution on straps of constant amplitude and three different toroidal phasing. A poloidal phasing of $\pi/2$ is imposed for load resilience. The toroidal phasing (0,2.9,3.8,0.4) minimize the edge LH power losses.}
    \label{fig:ITER_min}
\end{figure}

The ITER antenna is composed of 24 straps grouped into triplets (\cite{LAMALLE2013517}). For a fixed and even current amplitude on the straps, the remaining degree of freedom left to minimize the power losses is to change the phase distribution of the array. The spectrum minimizing the edge power losses found with ANTITER IV corresponds to the phasing (0,2.9,3.8,0.4) and is presented in figure \ref{fig:ITER_min}a along with the conventional phasings $(0\pi\pi0)$ and $(0\pi0\pi)$. Figure \ref{fig:ITER_min}b presents the related edge power loss spectrum. The respective percentage of power lost $P_{loss}/P_{tot}$ for each phasing is 0.25, 0.36 and 1.03 \%. While those numbers are small, the power coupled to the plasma is in the MW range (10 MW for one ITER launcher) which leads to power losses of about 10 kW (100 kW in ITER).

\begin{figure}
    \centering
    \includegraphics[width=\linewidth]{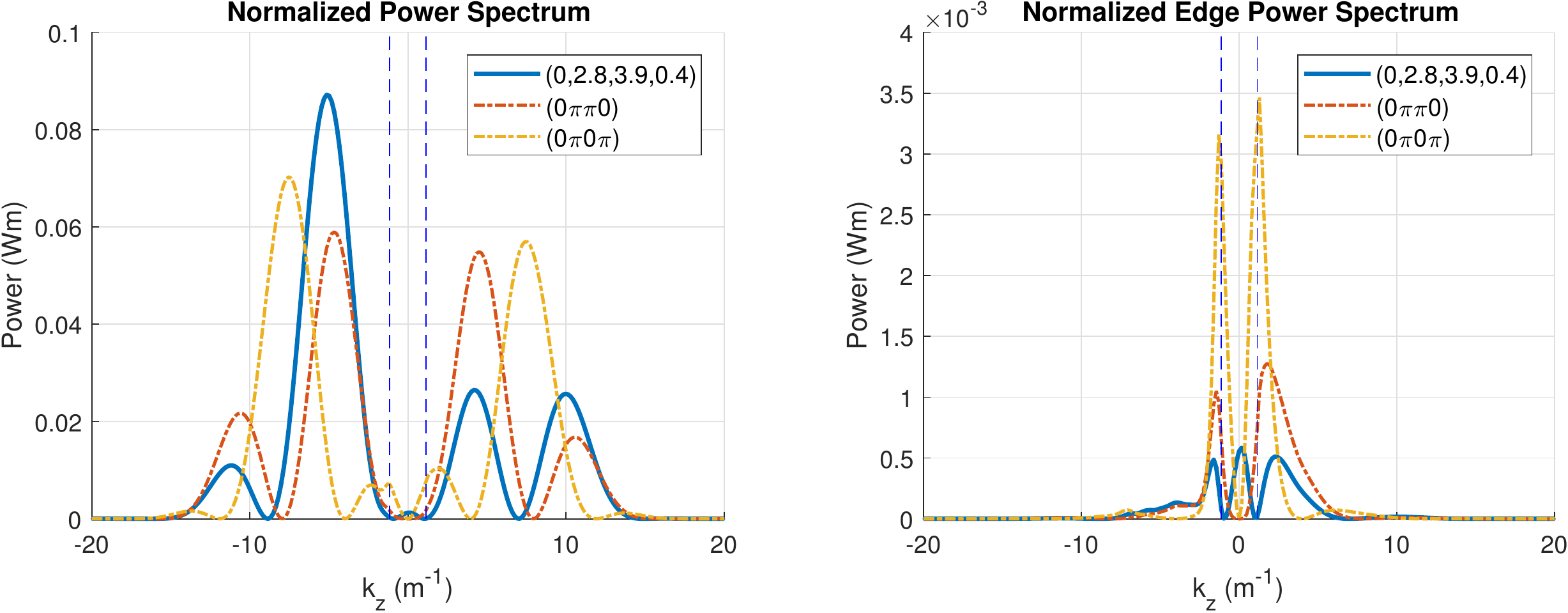}
    \caption{(a) Normalized $k_z$ power spectrum and (b) normalized $k_z$ edge power loss spectrum for a current distribution on straps of constant amplitude and three different toroidal phasing. A poloidal phasing of $\pi/2$ is imposed for load resilience. The toroidal phasing (0,2.8,3.9,0.4) minimize the edge LH power losses.}
    \label{fig:ITER_min_tilted}
\end{figure}

A misaligned antenna box, but with aligned FS, deforms the power loss spectrum but does not lead to direct spurious $E_z$ excitation and should only modestly change the result above. The same computation as in figure \ref{fig:ITER_min} but for a magnetic field tilted at an angle of 15$^\circ$ is presented in figure \ref{fig:ITER_min_tilted}. It leads to the phasing (0,2.8,3.9,0.4) and a respective percentage of power loss $P_{loss}/P_{tot}$ of 0.55, 0.95 and 1.61 \%. 

\begin{figure}
    \centering
    \includegraphics[width=\linewidth]{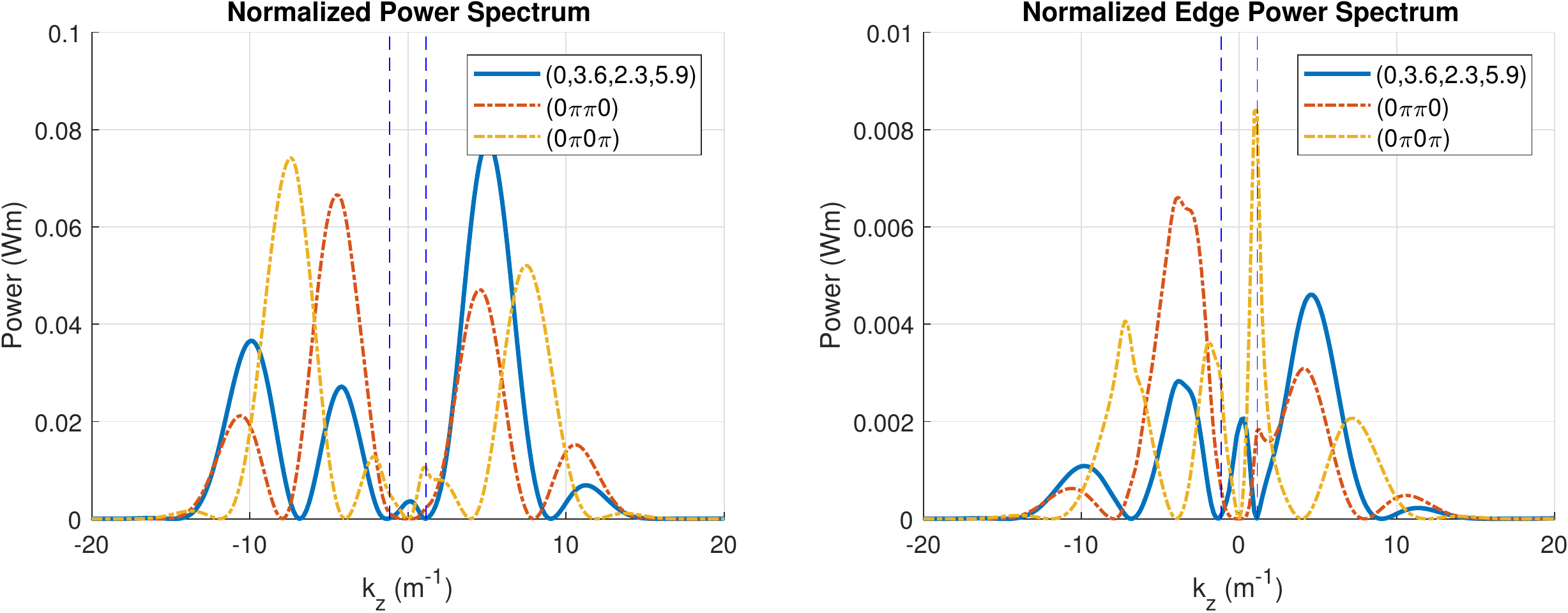}
    \caption{(a) Normalized $k_z$ power spectrum and (b) normalized $k_z$ edge power loss spectrum for a current distribution on straps of constant amplitude and three different toroidal phasing. A poloidal phasing of $\pi/2$ is imposed for load resilience. The toroidal phasing (0,3.6,2.3,5.9) minimize the edge LH power losses.}
    \label{fig:ITER_min_tilted_misaligned}
\end{figure}

While for the field aligned FS case the minimization only leads to a marginal decrease of the power loss at the LH resonance, a non-aligned FS will create an undesirable $E_z$ excitation and will greatly increase the power loss at the LH resonance. The misalignment of the FS with the magnetic field can be treated in ANTITER IV using the poloidal electric field $E_y$ computed in the aligned case and rotating it by an angle of 15$^\circ$. The result is presented in figure \ref{fig:ITER_min_tilted_misaligned} and leads to the phasing (0,3.6,2.3,5.9) and a respective percentage of power loss $P_{loss}/P_{tot}$ of 6.48, 7.98 and 7.12 \%.

\begin{figure}
    \centering
    \includegraphics[width=\linewidth]{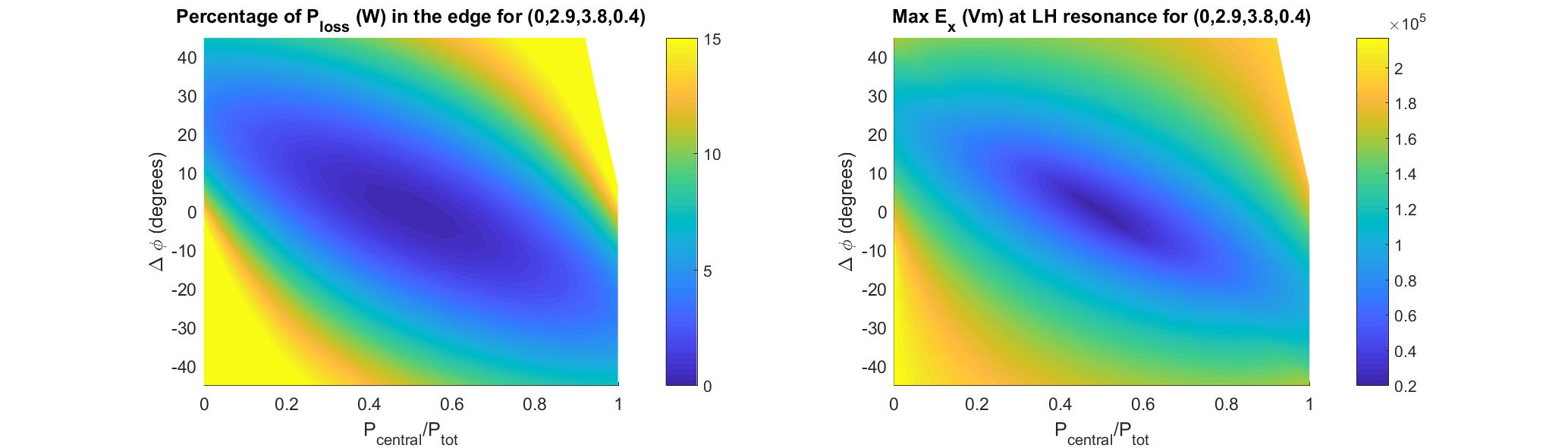}
    \caption{Map showing respectively: (a) the \% of power loss in the edge and (b) The maximum electric field at the LH resonance as a function of the power ratio $P_{central}/P_{tot}$ and a central phase deviation $\Delta \phi$ from the phasing minimizing the edge power loss $(0,2.9+\Delta \phi,3.8+\Delta \phi,0.4)$.}
    \label{fig:min_Ex}
\end{figure}

One can finally verify that minimizing the power loss at the lower hybrid corresponds to the minimization of the local radial electric field $E_x$ at this resonance. This is performed with a FS and an aligned antenna box by toroidally varying the power ratio between the two inner straps and the total power coupled $P_{central}/P_{tot}$ and by adding a phase $\Delta \phi$ to the best phasing $(0,2.9+\Delta \phi,3.8+\Delta \phi,0.4)$. The result, displayed in figure \ref{fig:min_Ex}, corresponds indeed to a minimum of $E_x$ excitation.

\section{Conclusion}
\label{sec:conclusion}
The paper presents an analytical derivation of the power loss that arises in the presence of a lower hybrid resonance in the plasma edge of a fusion machine. To do so we used a slab geometry and a cold plasma model. The power loss found is linked to the local radial electric field $E_x$ at the LH resonance position and is \textit{inversely} proportional to the slope along the radial direction $x$ of the first cold dielectric tensor component $\e_1$. The analytical formula of the power loss is then used to verify the accuracy of the numerical integration performed by the semi-analytical code ANTITER IV using the same slab description and cold plasma model. Good agreement is found between the two. Finally, we explore possible scenarios that could minimize the ITER ICRH power losses: 
\begin{enumerate}
    \item In a screen-aligned scenario, one should avoid the excitation of the lower $\abs{k_z}<k_0$ part of the antenna power spectrum.
    \item In case of direct $E_z$ excitation due to a misalignment of the FS with the background magnetic field, the lower $k_z$ region to be avoided in the power spectrum is enlarged above $k_0$.
\end{enumerate}
The fact that the power loss is proportional to the derivative of $\e_1$ along $x$ provides a method to directly influence the power loss at the LH resonance by shaping the plasma density profile near the antenna. An easy way to do so would be to use gas puff (\cite{ZHANG2019364}) and will be explored in a future paper.

While the results are in line with earlier work (\textit{e.g.} \cite{morales,Lawson_1992}), the limits of the model should also be emphasized. The model neglects finite temperature effects preventing the detailed description of the wave conversion at the LH resonance to new electrostatic waves (\textit{e.g.} ion Berstein waves). The model does not take into account the poloidal and toroidal inhomogeneity of the plasma density profile. It also neglects possible non-linear effects (\textit{e.g.} ponderomotive force) arising in the presence of strong fields excited by the antenna. The model also uses plane geometry and an antenna recessed into the wall of the machine.

\section*{Acknowledgements}
This work has been carried out within the framework of the EUROfusion Consortium and has received funding from the Euratom research and training programme 2014-2018 and 2019-2020 under grant agreement No 633053. The views and opinions expressed herein do not necessarily reflect those of the European Commission.


\section*{Declaration of interests}
The authors report no conflict of interest.

\appendix
\section{A theorem about matrix differential equations}\label{app:thm}
\begin{theorem}
Let be a first-order matrix ordinary differential equation of the form
\begin{align}
    \dv{t}\mathbf{{x}}(t) = \mathbf{A}(t)\mathbf{x}(t)
\end{align}
with $\vb{x}$ a $n \times 1$ vector and $\vb{A}$ a $n\times n$ matrix. 
If $\vb{A}(t)$ commutes with its integral $\int^t \vb{A}(s) \dd s$ then the general solution of the differential equation is 
\begin{align}
    \mathbf{x}(t)=e^{\int^t \mathbf{A}(s) \dd s} \mathbf{c},\label{sol}
\end{align}
where $\vb{c}$ is an $n \times 1$ constant vector.
\label{th:1}
\end{theorem}\

\begin{proof}
{\small
Using the definition of the matrix exponential, the solution \eqref{sol} can be written
\begin{align}
\mathbf x(t) &= e^{\int^t \mathbf{A}(s) \dd s} \mathbf{c}=\sum_{n=0}^\infty \frac{\left(\int^t  \mathbf{A}(s) \dd s\right)^n}{n!} \mathbf{c}.
\end{align}
Its derivative reads
\begin{align}
\dv{t}\mathbf{x}(t) &= \dv{}{t}\left(\sum_{n=0}^{+\infty} \frac{\left(\int^t  \mathbf{A}(s) ds\right)^n}{n!} \right)  \mathbf{c}, \\
&= \mathbf{A}(t) \sum_{n=0}^{+\infty}  \frac{\left(\int^t  \mathbf{A}(s) ds\right)^{n-1}}{(n-1)!}\mathbf{c}\\
&=\mathbf A(t) e^{\int^t\mathbf A(s)\dd s},
\end{align}
where the second equality follows from the fact that, if $\int^t\mathbf A(s)\dd s$ commutes with its $t$ derivative $\mathbf A(t)$, one can write
\begin{align}
    \dv{t}\left(\int^t  \mathbf{A}(s) ds\right)^n=n\mathbf A(t)\left(\int^t  \mathbf{A}(s) ds\right)^{n-1}.
\end{align}
We finally have
\begin{align}
\dot{\mathbf x}(t) &=  \mathbf{A}(t) \mathbf x(t).
\end{align}}
\end{proof}

\section{Proof of Equation \eqref{eq:alpha}}\label{app:alpha}
We will here provide a proof of the identity
\begin{align}
    \Im[\alpha(x)]\eval_{x=-\kappa}^{x=+\kappa}\simeq-\frac{\pi}{k_0^2\abs{\epsilon'_1}}.
\end{align}
To prove this assertion, one has to notice that, in fact, the LH resonance is not exactly located at $x=0$. Because of the collisions arising in the plasma, the frequency $\omega$ is not real but possesses a small, positive, imaginary part:
\begin{equation}
    \omega=\Re\omega+i\Im\omega,\qquad \abs{\Im\omega}\ll\abs{\Re\omega}\text{ and }\Im\omega>0.
\end{equation}
Consequently, $\epsilon_1$ can be expanded as
\begin{align}
    \epsilon_1(\omega)=\epsilon_1(\Re\omega+i\Im\omega)=\epsilon_1(\Re\omega)+i\Im\omega\frac{\partial\epsilon_1}{\partial\omega}\eval_{\omega=\Re\omega}+\mathcal O\qty((\Im\omega)^2)
\end{align}
and also exhibits a small imaginary part, $\Im\epsilon_1\simeq\Im\omega\frac{\partial\epsilon_1}{\partial\omega}$. Recalling ourselves that $\epsilon_1(\omega)=1-\sum_\alpha\frac{\omega_{p\alpha}^2}{\omega^2-\omega_{c\alpha}^2}$, one can show that $\Im\epsilon_1$ is indeed positive:
\begin{align}
    \Im\epsilon_1\simeq\Im\omega\frac{\partial\epsilon_1}{\partial\omega}\simeq2\Re\omega\Im\omega\frac{\partial\epsilon_1}{\partial\omega^2}\simeq2\Re\omega\Im\omega\sum_\alpha\frac{\omega_{p\alpha}^2}{\qty[\qty(\Re\omega)^2-\omega_{c\alpha}^2]^2}>0.
\end{align}
In the following, we will simply write $\omega$ instead of $\Re\omega$. The main consequence of the discussion above is that $\epsilon_1(x)$ doesn't vanish anymore at $x=0$, but at $x=\bar x\equiv-\frac{i\Im\epsilon_1}{\epsilon'_1}$. In other words, the LH resonance is not anymore located at $x=0$, but at $x-\bar x=0$. This fact is easily implemented in \eqref{eq:pre_DP} by replacing  $\Im\alpha(x)\eval_{x=-\kappa}^{x=+\kappa}$ by
\begin{align}
 \Im\alpha(x)\eval_{x-\bar x=-\kappa}^{x-\bar x=+\kappa}&=\frac{1}{k_0^2}\Im\qty[\frac{\log(\kappa+\bar x)-\log(-\kappa-\bar x)}{\epsilon'_1}]\\
 &\simeq\frac{1}{k_0^2}\frac{\arg\qty(\kappa+\bar x)-\arg\qty(-\kappa-\bar x)}{\epsilon'_1}.\label{eq:int}
\end{align}
Noticing that, for $\abs{\kappa}\gg\abs{\frac{\Im\epsilon_1}{\epsilon'_1}}$, one has
\begin{align}
    \arg\qty(\kappa+\bar x)&\simeq 0,\\
    \arg\qty(-\kappa-\bar x)&\simeq
    \left\lbrace\begin{array}{ll}
        \pi, & \epsilon'_1>0 \\
        -\pi, & \epsilon'_1<0
    \end{array}\right..
\end{align}
Equation \eqref{eq:int} becomes
\begin{align}
     \Im\alpha(x)\eval_{x-\bar x=-\kappa}^{x-\bar x=+\kappa}\simeq-\frac{\pi}{k_0^2\abs{\epsilon'_1}},
\end{align}
which is the desired result.

\end{document}